\documentclass[manuscript,screen]{acmart}

\AtBeginDocument{%
  }

\setcopyright{acmlicensed}
\copyrightyear{2018}
\acmYear{2018}
\acmDOI{XXXXXXX.XXXXXXX}

\acmConference[Conference acronym 'XX]{Make sure to enter the correct
  conference title from your rights confirmation emai}{June 03--05,
  2018}{Woodstock, NY}
\acmISBN{978-1-4503-XXXX-X/18/06}

\definecolor{TODOcolor}{HTML}{fa9fb5}

\definecolor{cormancolor}{HTML}{FF0000}

\definecolor{fancolor}{HTML}{dd3497}

\definecolor{editcolor}{HTML}{0570b0}

\definecolor{arlencolor}{HTML}{FF0000}

\newcommand{\addition}[0]{}

\newcommand{\minor}[0]{}

\begin{document}

\title[A Framework for Combating Misinformation]{Skeptik: A Hybrid Framework for Combating Potential Misinformation in Journalism}

\author{Arlen Fan}
\authornote{Both authors contributed equally to this research.}
\email{afan5@asu.edu}
\author{Fan Lei}
\authornotemark[1]
\email{flei5@asu.edu}
\author{Steven R. Corman}
\email{steve.corman@asu.edu}
\author{Ross Maciejewski}
\email{rmacieje@asu.edu}
\affiliation{%
  \institution{Arizona State University}
  \city{Tempe}
  \state{Arizona}
  \country{USA}
}

\renewcommand{\shortauthors}{Fan and Lei, et al.}

\begin{abstract}
The proliferation of misinformation in journalism, often stemming from flawed reasoning and logical fallacies, poses significant challenges to public understanding and trust in news media. Traditional fact-checking methods, while valuable, are insufficient for detecting the subtle logical inconsistencies that can mislead readers within seemingly factual content. To address this gap, we introduce Skeptik, a hybrid framework that integrates Large Language Models (LLMs) with heuristic approaches to analyze and annotate potential logical fallacies and reasoning errors in online news articles. Operating as a web browser extension, Skeptik automatically highlights sentences that may contain logical fallacies, provides detailed explanations, and offers multi-layered interventions to help readers critically assess the information presented. The system is designed to be extensible, accommodating a wide range of fallacy types and adapting to evolving misinformation tactics. Through comprehensive case studies, quantitative analyses, usability experiments, and expert evaluations, we demonstrate the effectiveness of Skeptik in enhancing readers' critical examination of news content and promoting media literacy. Our contributions include the development of an expandable classification system for logical fallacies, the innovative integration of LLMs for real-time analysis and annotation, and the creation of an interactive user interface that fosters user engagement \addition{and close reading}. By emphasizing the logical integrity of textual content rather than relying solely on factual accuracy, Skeptik offers a comprehensive solution to combat \addition{potential} misinformation in journalism. Ultimately, our framework aims to \addition{improve critical reading and} protect the public from deceptive information online and enhance the overall credibility of news media.
\end{abstract}




\maketitle
\section{Introduction}
\label{section:introduction}

The rapid dissemination of news serves as a critical resource across various domains, from staying up to date with global events to making well-informed decisions in dynamic sectors like the stock market. Journalists strive to present information in a way that is accessible and comprehensible to a broad audience. However, the complexity of certain topics can lead to oversimplifications or the use of persuasive techniques that may mislead readers.

While journalists aim to inform the public, there is a growing concern about the use of flawed reasoning and logical fallacies in news content. These fallacies can make arguments seem valid when they are not, \addition{potentially} misleading readers and distorting public perception. \addition{For example, a false causality fallacy can appear in statements like, ``\textit{Extreme weather-related deaths in the U.S. \textbf{have decreased} by more than 98\% over the last 100 years; therefore, global warming \textbf{saves lives}.}'' Here, a correlation is incorrectly presented as causation, creating a misleading impression.} This issue is pervasive in text content, making it important to examine articles for potential misinformation that stems from faulty logic~\cite{wardle2017information}.

\addition{Recent studies have highlighted the prevalence of logical fallacies in misinformation. For instance, Zanartu et al.~\cite{zanartu2024technocognitive} developed a dataset mapping climate misinformation to reasoning fallacies, demonstrating that such fallacies are common mechanisms for creating misinformation. Similarly, Jin et al.~\cite{jin-etal-2022-logical} introduced a dataset for logical fallacy detection, further emphasizing the frequency of fallacious reasoning in misleading content. These findings suggest that logical fallacies are not isolated incidents but are widespread across various domains of misinformation.}

Researchers have developed models to pinpoint and analyze the spread of misinformation, examining dissemination patterns on social networks~\cite{ma2018rumor, vosoughi2018spread}, stylistic writing cues~\cite{bond2017lyin, pisarevskaya2017rhetorical}, and the credibility of sources~\cite{castillo2011information, viviani2017credibility}. These models leverage textual information and their context to identify fake news. But combating misinformation remains challenging. Most tools aimed at fighting misinformation involve fact-checking by professionals or laypeople, either manually or with the help of automated means for scalability and speed. They also provide credibility indicators for content~\cite{zhang2018structured}. However, fact-checking alone is not enough\addition{, as narratives may employ factually correct data, but apply logical fallacies to mislead the audience to a desired conclusion}. \addition{Such} logical fallacies can be harder to identify. Some tools~\cite{zavolokina2024think, jahanbakhsh2022our, jahanbakhsh2024browser, chakraborty2016stop, rony2018baitbuster} analyze the validity of arguments and assess the coherence of narratives. Our proposed tool seeks to analyze the argumentative coherence of the article as a whole, an underexplored way of combating misinformation.

News websites may use fallacies to construct arguments which, although seemingly valid, are based on flawed reasoning or appeal to common cognitive biases. This issue is made worse by instances where opinion pieces are misrepresented as factual news, a problem that even reputable outlets have encountered~\cite{rodrigo2023systematic}. This situation highlights a critical limitation of current fact-checking tools: their inability to discern the overarching validity of an argument composed of purportedly factual statements. Our approach seeks to address this challenge by focusing on the logical integrity of textual information, ensuring that articles do not exhibit faulty reasoning. Recent breakthroughs in Large Language Models (LLMs) offer promising capabilities in pattern recognition~\cite{brown2020language}, real-time analysis~\cite{chen2023real}, and even educational potential~\cite{moore2023empowering}, presenting new techniques for addressing these challenges.

As such, addressing misinformation in journalism necessitates a holistic approach that considers the textual content of articles. This is the aim of our research. \minor{In this work, we ask: \textit{How can LLMs be integrated into an interactive framework to assist users in identifying and critically reflecting on logical fallacies in online news?} We answer this question through the design, implementation, and evaluation of a system that embeds LLMs within a human-in-the-loop framework for fallacy detection and annotation.} While existing tools have made strides in detecting textual misinformation via AI-based methods~\cite{nguyen2018believe, wang2022rumorlens, trokhymovych2021wikicheck}, a significant gap remains in analyzing the logical coherence of arguments within articles. The subtleties of misinformation, along with the complexity of logical fallacies, underscore the need for a tool that can validate the soundness of arguments and educate the audience on how to critically assess news in a real-world setting. Our work aims to bridge this gap and support readers in critically examining information.

In this paper, we introduce a framework with a web browser extension that annotates online news for potential logical fallacies and reasoning errors. Our approach transcends traditional fact-checking by focusing on the logical integrity of the arguments presented, significantly impacting the message credibility of the information. This analysis is important for providing readers with a holistic understanding of the narrative, a key aspect of media literacy~\cite{livingstone2004media}, whose importance is increasing with the continuing growth of misinformation~\cite{lord2021strengthen, zuckerman2021study}. This end-to-end solution aims to protect the public from misinformation online. \addition{Skeptik is designed with modularity in mind, allowing future extensions that incorporate established fact-checking methodologies to detect a broader range of misinformation. However, it is crucial to acknowledge that LLMs, while powerful, are not infallible. They can sometimes produce outputs that are convincingly wrong or reflect inherent biases~\cite{lin-etal-2022-truthfulqa}. Therefore, Skeptik is designed not as a definitive authority but as a utility tool to aid users in critically evaluating content. It flags possible fallacies with language such as "This may be an example of...", leaving the final judgment to the user's discretion.} Through case studies and a series of quantitative and human evaluation methods, we demonstrate the effectiveness of our tool in flagging potentially deceptive information.

In summary, our contributions include:
\begin{itemize}
\item Development of a framework that integrates LLMs with heuristic approaches for analyzing \addition{potential} logical fallacies and reasoning errors in journalism;
\item Introduction of a web browser extension that annotates online news with an analysis of logical fallacies;
\item Case studies, a quantitative study, usability experiments, and expert evaluations of this tool, demonstrating the effectiveness of the framework in identifying and annotating \addition{potentially} misleading information.
\end{itemize}

\section{Related Work}
\label{sec:related_work}

\addition{Combating misinformation in news articles is a complex societal problem that requires systematic and interdisciplinary solutions combining computer science technologies and psychology.} In this section, we summarize related works on fake news detection, logical fallacy detection and intervention, and narrative visualizations.

\subsection{Fake News Detection}
\label{subsection:fake_news_detection}

\addition{In the study of fake news detection, researchers have primarily focused on a range of methodologies, including fact-checking, decision-support systems, and computational techniques.} Fake news is often categorized as misinformation in the guise of authentic news, highlighting the challenges in discerning truth from falsehood~\cite{wardle2017information}. Traditional detection methods primarily focus on the analysis of news content, including lexicon, syntax, semantics, and discourse levels, as well as on how fake news propagates on social networks~\cite{zhou2019fake}. To enhance detection capabilities, researchers have integrated diverse features, including author profiles~\cite{zhang2020fakedetector}, social media analytics~\cite{karduni2019vulnerable}, and the dynamics of news propagation~\cite{wu2015false, vosoughi2018spread}. These advancements underscore the complexity of misinformation, which not only manifests through factual inaccuracies but also through the manipulation of logical structures within seemingly valid narratives. RumorLens~\cite{wang2022rumorlens} used a combination of NLP, propagation networks, and features about the author to analyze false information on social media. Nguyen et al.~\cite{nguyen2018believe} proposed a UI system that leverages automated algorithms for information retrieval and machine learning, along with interactive elements that allow users to apply their own critical thinking and knowledge to the verification process. Their findings suggest that such hybrid approaches can improve the accuracy of fact-checking by involving users in the evaluation process, thus addressing some of the scalability issues inherent in manual fact-checking methods while also mitigating the risks of over-reliance on automated systems. \addition{Beyond these tools, comprehensive surveys have been conducted to map the landscape of automated fact-checking. Guo et al.~\cite{10.1162/tacl_a_00454} provided an extensive overview of the field, discussing various datasets, models, and the integration of natural language processing techniques in fact-checking pipelines. Their work highlights the challenges in automating fact-checking processes and underscores the need for systems that can handle diverse information sources.}

Despite these advancements, existing end-to-end solutions primarily focus on factual verification through evidence retrieval and NLI classification. 
\minor{One such system is ClaimBuster~\cite{hassan2017claimbuster}, which,
however, relies on closed external components and has not been tested against benchmark datasets. Another recent work by Chernyavskiy et al.~\cite{chernyavskiy2021whatthewikifact} proposed a high-level solution with a more comprehensive conceptual pipeline. While these systems have contributed to the field, they have limitations in terms of openness and benchmark evaluation.}
Our study extends beyond traditional fact-checking and fake news detection by introducing a method that analyzes not just the factual accuracy but also the logical coherence of news content. We recognize that fallacies can mask misinformation within factually accurate statements, thus, our research aims to uncover these potentially deceptive tactics. By doing so, we are able to offer another approach for combating potential misinformation.

\subsection{\addition{Combating Logical Fallacies}}
\label{subsection:combating_logical_fallacy}

\addition{Psychological research~\cite{schmid2019effective} has identified that combating misinformation requires both fact-based (e.g., factual explanation) and technique-based (e.g., detection and intervention of logical fallacies) approaches. However, compared to fact-checking works, logical fallacy detection and intervention are still in an early stage of interdisciplinary research that involves the synthesis of psychological and computer science~\cite{zanartu2024technocognitive}.}

\addition{\textbf{Logical Fallacy Detection.} Previous research has explored computational methods for detecting logical fallacies across various contexts. Habernal et al.~\cite{habernal-etal-2017-argotario} developed Argotario, a gamified platform aimed at identifying fallacies within dialogues. Wachsmuth et al.~\cite{wachsmuth2017computational} examined the quality dimensions of arguments, including logical sufficiency. Additionally, Sahai et al.~\cite{sahai2021breaking} compiled a dataset of Reddit discussions annotated for multiple fallacy types, facilitating the study of fallacy detection in online discourse. More closely related to our work, Musi et al.~\cite{musi2022developing} analyzed a corpus of fact-checked COVID-19 news to identify ten logical fallacies that systematically trigger misinformation. These studies were confined to single datasets, which limits their generalizability across diverse domains—an essential consideration for effective fallacy detection in real-world scenarios. More recent work improves the fallacy detection performance by using multiple datasets and LLMs. Li et al.~\cite{li-etal-2024-reason} presented a structured evaluation framework to systematically measure and enhance LLMs' understanding of logical fallacies, while Jeong et al.~\cite{jeong2025large} addressed the limitations of current LLMs in logical fallacy detection by proposing a novel prompting strategy enriched with implicit contextual information.
Currently, the primary focus of technical research for logical fallacy detection lies within the domain of natural language processing (NLP). However, identifying logical fallacies constitutes only the first step in mitigating their impact on individuals; it is equally important to explore effective intervention strategies for reducing their influence.}

\addition{\textbf{Logical Fallacy Intervention.} Intervening against logical fallacies is essential for mitigating the spread of misinformation and enhancing critical thinking. Recent research has explored various strategies to address this challenge, ranging from technological solutions to psychological interventions. Ecker et al.~\cite{ecker2022psychological} examined the psychological underpinnings of misinformation belief and its resistance to correction. They highlighted factors like cognitive biases and social dynamics that contribute to the persistence of false beliefs. Their review emphasized the effectiveness of both preemptive (``prebunking'') and reactive (``debunking'') interventions, suggesting that understanding these psychological drivers is crucial for designing effective misinformation countermeasures. Addressing the role of cognitive biases in information processing, Draws et al.~\cite{Draws_Rieger_Inel_Gadiraju_Tintarev_2021} proposed a checklist adapted from business psychology to combat biases in crowdsourcing tasks. This checklist serves as a practical tool to identify and mitigate biases that may affect data quality. In the educational domain, \minor{Hruschka and Appel~\cite{hruschka2023learning}} conducted an experimental intervention to assess the impact of teaching informal fallacies on fake news detection. This study underscores the potential of educational interventions in enhancing critical thinking skills related to misinformation.
Zanartu et al.~\cite{zanartu2024technocognitive} discussed the fallacies detection and intervention methods in climate misinformation by combining psychological insights with computational methods. Robbemond et al.~\cite{10.1145/3503252.3531311} investigated how different explanation modalities affect users' ability to assess information credibility. Their findings indicate that explanations can improve users' understanding and trust in AI systems, thereby enhancing the effectiveness of misinformation mitigation.}

\addition{While existing studies treat detection and user intervention as separate processes, there remains a significant need for a unified solution that seamlessly integrates these elements. By merging detection and intervention into a cohesive workflow, our work addresses a critical gap between current human-computer interaction (HCI) solutions and machine learning-based computational methods, offering a novel engineering solution to combat misinformation in the consumption of digital news.}

\subsection{\addition{Narrative Visualization}}
\label{subsection:narrative_vis}

\addition{The design space of narrative visualization has been extensively explored within the visualization community, resulting in a diverse array of storytelling and annotation techniques aimed at highlighting insights and effectively conveying core messages to audiences~\cite{Segel2010, Hullman2011, Tong2018}. These techniques utilize structured narratives and annotations to enhance the clarity, engagement, and effectiveness of visual data representations.}
\addition{Various visualization tools have been developed that incorporate advanced narrative techniques. For instance, Chen et al.~\cite{ren2017chartaccent} developed interactive annotation systems tailored to facilitate clearer communication of critical patterns within data visualizations. Satyanarayan et al.~\cite{Satyanarayan2014} have also contributed tools that automate the creation of presentation-style storytelling visualizations, improving the efficiency and consistency of narrative delivery. Recent research has leveraged advances in deep learning to further automate the generation of annotations, integrating sophisticated feature extraction techniques with natural language generation frameworks~\cite{Lai2020, Liu2020}. These innovations enhance the capacity of narrative visualization tools to automatically detect and describe important patterns and trends in data, thereby enriching the storytelling experience.} \addition{Additionally, several systems explicitly focus on annotations to support data-driven storytelling~\cite{Srinivasan2019, Chen2020, choudhry2020once, lei2023geoexplainer}. Kosara and Mackinlay~\cite{kosara2013storytelling} particularly emphasize the crucial role annotations play in visualization storytelling, noting their value in highlighting key insights and guiding audience interpretation.}
\addition{Building on advances in narrative visualization and interactive annotation techniques, our framework automatically annotates detected logical fallacies and establishes visual linkages to multi-layered narrative interventions, enhancing understanding and facilitating critical thinking.}

\section{An Expandable Framework for Detecting and Annotating Fallacies}
\label{section:expandable_framework}

\begin{figure}[t]
\centering	
\includegraphics[width=0.7\linewidth]{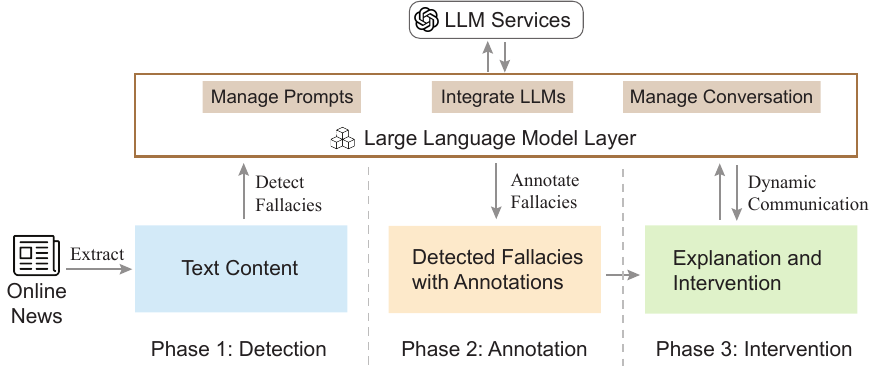}
\caption{Pipeline of our framework. We extract text elements from news and return annotations about detected fallacies. Readers are able to interact dynamically with the article and the annotations by communicating with the large language model module.}
\Description{(Flowchart) The figure shows the framework pipeline that contains three major phases. Three phases include: 1) extracting article content and detecting potential fallacies, 2) annotating the detected fallacies, and 3) providing interventions to reduce the influence of fallacies.}
\label{fig:framework-pipeline}
\end{figure}

In this section, we introduce our annotation framework designed to highlight parts of articles that may contain potential misinformation and explain the potentially misleading items. Our technique stands apart from prior research works~\cite{nguyen2018believe,trokhymovych2021wikicheck} and existing platforms such as Snopes or PolitiFact, which primarily serve as fact-checking resources. Our approach has three phases (Figure~\ref{fig:framework-pipeline}) that use large language models (LLMs) for dynamically providing 1) detection, 2) annotation, and 3) intervention to logical fallacies, enabling a more thorough examination of narratives.

We first describe inoculation, a concept in communication theory that inspired this framework. We then introduce our design objectives. Next, we explain our methods for extracting content from online news sources and describe our expandable classification system for logical fallacies. Finally, we introduce the design of the Large Language Model Layer, which handles the interface communication.

\subsection{Inoculation: Inspiration for this Tool}
\addition{Our initial concept for this tool was informed by a series of formative interviews conducted with a communication expert, who is also a co-author of this paper. These interviews were structured as semi-formal discussions aimed at exploring theoretical frameworks that could underpin the framework's design. Over the course of three sessions, each lasting approximately 60 minutes, we delved into various communication theories, with a particular focus on inoculation theory as articulated by Pfau and Burgoon~\cite{pfau1988inoculation}. This theory posits that exposing individuals to weakened forms of persuasive arguments can bolster their resistance to subsequent, stronger persuasive attempts.}

\addition{During these sessions, we examined how the principles of inoculation—such as introducing a manageable threat to beliefs and providing counterarguments—could be operationalized within a digital tool aimed at identifying and annotating logical fallacies in online news content. The expert provided insights into how these theoretical concepts could translate into practical features, such as highlighting specific fallacies and offering users strategies for rebuttal.} \addition{Additionally, the expert provided concrete examples of common persuasive strategies and logical fallacies frequently encountered in online news articles, which helped inform the selection of fallacy types that our tool focuses on detecting and annotating.} \addition{This collaborative exploration established the foundation for our framework, aimed at strengthening users' critical thinking skills and resilience to misinformation through the principles of inoculation. Building on this theoretical groundwork, we developed a system to detect and annotate potential fallacies in online news. The key components of our approach are outlined in the following sections.}

\subsection{Design Objectives for Measurable Evaluation}
\label{subsection:design_obj}

To ensure the practicality and effectiveness of our proposed framework in combating misinformation, we have identified specific design objectives \addition{through a collaborative process involving a series of co-design workshops and feedback sessions with the communication expert. In these sessions, we presented early prototypes and gathered feedback on aligning the tool's functionalities with principles from communication theory.} These objectives guide the development process (Sections~3.2-3.5 and~\ref{section:interface-design}) and provide measurable criteria for evaluating (Section~\ref{section:evaluation}) the framework's performance. We distinguish between system design objectives and UI design objectives as follows:

\textbf{System Design Objectives:}
\begin{itemize}
    \item \textbf{SDO1: Identifying potential fallacies in online news automatically.}
To combat misinformation, it is important to identify embedded fallacies without relying heavily on manual efforts or specialized domain knowledge. Traditional methods require domain experts to manually weigh in with their knowledge, which makes these efforts expensive. Our tool automates the process, leveraging the capabilities of language models to identify fallacies~\cite{li-etal-2024-reason, jeong2025large}. \addition{Recognizing that LLMs are rapidly evolving, the system should be modular and flexible, allowing it to easily incorporate more advanced models and improved prompt strategies as they emerge.}

    \item \textbf{SDO2: Annotating detected fallacies (and other relevant information) dynamically.}
Once a potential fallacy is identified, it is important to communicate this information properly to the reader. Therefore, our tool will dynamically annotate or mark sentences within an article that correspond to a detected fallacy. This feature visually guides readers through the content, which in turn enables a more critical engagement with the material. Additionally, we will address issues related to the coherence between charts and text, ensuring that any discrepancies or misleading representations are clearly indicated. \addition{Here, we note that LLMs do have the potential to hallucinate or misidentify a sentence as a logical fallacy. By presenting these as potential fallacies, our system encourages close reading and critical thinking.}

    \item \textbf{SDO3: Adapting to varying fallacy types.}
Given the variety of fallacies that can be employed to distort information, our tool will be flexible to adapt to a wide range of identified fallacies. The classification of these fallacies can vary: Musi and Reed~\cite{musi2022fallacies} outline 10 distinct categories of fallacies, whereas Lisnic et al.~\cite{lisnic2023misleading} have identified 7 specific reasoning errors in the context of data visualization. In the domain of argument appraisal, Tindale's work~\cite{tindale2007fallacies} lists an extensive 47 fallacies. Our initial framework will incorporate the fallacies as classified by Musi and Reed~\cite{musi2022fallacies} and Lisnic et al.~\cite{lisnic2023misleading}, but it is engineered to be extensible, allowing for the integration of additional fallacies.
\end{itemize}

\textbf{UI Design Objectives:}
\begin{itemize}
    \item \textbf{UIDO1: Providing
interventions to effectively reduce the fallacy influence.}
Beyond identifying (\textbf{SDO1}) and annotating (\textbf{SDO2}) fallacies, our tool will offer interventions designed to mitigate the influence of these fallacies on the reader. These interventions include providing clarifications, offering additional context, or presenting counterarguments to challenge misleading claims.  Guided by Cook et al.'s~\cite{cook2015misinformation} research on misinformation correction, our interventions \addition{are} designed in such a way to minimize the effect of these fallacies. These interventions are tiered and offer progressive clarification and contextual grounding to challenge potentially misleading arguments. Detailed elaboration of these intervention levels will be explained in Section~\ref{section:interface-design}.
\end{itemize}

\subsection{Extracting Content from Online News} To effectively detect potential fallacies from online news sources (\textbf{SDO1}), we need to recognize and extract the main contents and reformat them as input to the language modules.

\textbf{Extracting Text Content.} To automatically extract text content that adapts to the varying formats of news websites in real-time, inspired by Lin and Ho's work~\cite{10.1145/775047.775134}, our fully automatic approach is built upon ParEx~\cite{7745047} to collect the text content within an article while filtering out as much irrelevant information as possible. \addition{ParEx is a heuristic-based content extraction method that identifies meaningful article text by analyzing paragraph (<p>) tag structures in HTML pages. It clusters text blocks while filtering out unrelated web elements such as sidebars and ads, enabling lightweight and adaptable extraction across different news sites without requiring manual training or extensive customization.}

\begin{figure}[t]
\centering	
\includegraphics[width=\linewidth]{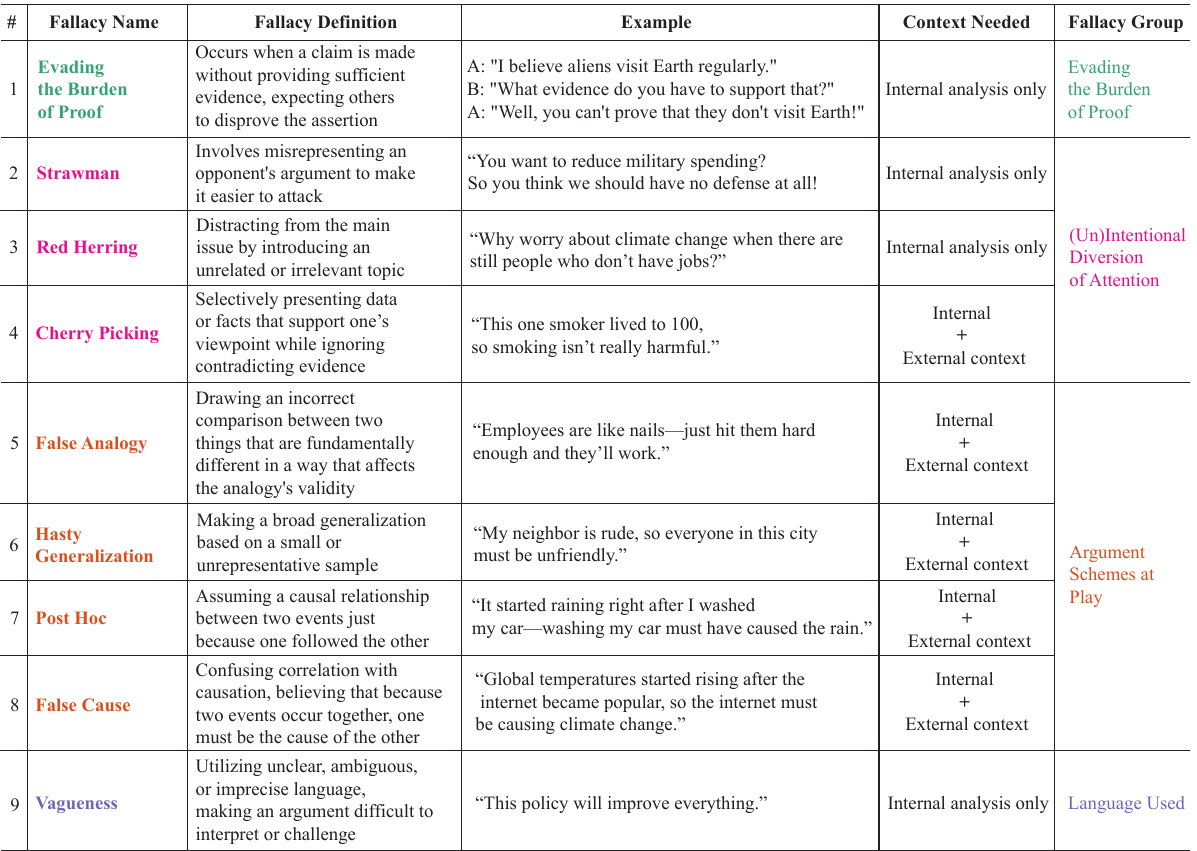}
\caption{The fallacies used in our framework \addition{with examples}, which are adapted from Musi and Reed~\cite{musi2022fallacies}. Fallacies are further grouped (far right column) and colored-coded accordingly~\cite{10.1179/000870403235002042}. \addition{The Context Needed column indicates whether each fallacy can be reliably identified based solely on the article’s content or if external information is necessary for accurate detection~\cite{zanartu2024technocognitive}.}
}
\Description{(Table) This table shows thirteen fallacy categories with grouped colors that are used in the framework.}
\label{fig:fallacy-table}
\end{figure}

\subsection{Classification for Logical Fallacies}
\label{subsection:classification_for_logical_fallacies}

Figure~\ref{fig:fallacy-table} outlines the classification of the fallacies used in our framework\addition{, along with an indication of whether each fallacy can be reliably identified based solely on the article’s content or if external information is necessary for accurate detection~\cite{zanartu2024technocognitive}.} \addition{Our selection primarily builds upon the comprehensive analysis by Musi and Reed~\cite{musi2022fallacies}, who systematically categorize fallacies commonly encountered in digital media, particularly those that frequently serve as misinformation triggers. From their extensive taxonomy, we strategically selected a subset of fallacies that were identified as especially prevalent in online news environments and influential in misinforming readers. This selection process was further refined through iterative discussions and evaluations conducted in close collaboration with a communication expert, ensuring practical relevance and analytical feasibility.}

\addition{While the chosen fallacies are not exhaustive compared to broader classifications such as Tindale's~\cite{tindale2007fallacies} expansive overview, our initial scope was intentionally constrained due to practical considerations and the current capabilities of LLMs in detecting and interpreting nuanced argumentative structures. However, the modular and expandable nature of our framework explicitly accommodates future enhancements. As advancements in LLM technology emerge, our system is designed to seamlessly integrate additional fallacies, thereby progressively expanding its analytical scope and aligning with our design objective \textbf{SDO3}.}

\subsection{Large Language Model Layer}
\label{subsection:prompt_engineering}

Our framework leverages the power of LLMs to combat fallacies in online news articles. The large language model layer serves as the middleware that handles the communication between the frontend interface and the LLMs. This layer supports three tasks as shown in Figure~\ref{fig:framework-pipeline}: 1) LLM prompt management, 2) flexible integration of LLMs, and 3) conversation management between the reader and the LLM.

\textbf{Prompt Management.} Within our prompt engine, we deploy tailored prompts specifically engineered to address the challenges of detecting logical fallacies. This approach leverages LLMs to parse and analyze entire arguments and assess logical soundness, distinguishing our methodology from traditional fact-checking mechanisms.

\textit{Detecting Logical Fallacies.} Our prompt is designed to identify an array of logical fallacies within the article text. Drawing upon the fallacy taxonomy classified by Musi and Reed~\cite{musi2022fallacies}, this prompt instructs the LLM to scan the article for indicative signs of each listed fallacy. The process entails:

\begin{itemize} \item Presenting the LLM with definitions and examples of each fallacy, thereby equipping the model with the necessary context for accurate identification. \item Requesting the LLM to output instances where these fallacies might be present, specifying the exact location (e.g., sentence numbers) and providing a rationale based on the input definitions. \end{itemize}

\noindent The full prompt is in Appendix~\ref{appendix:prompt-for-detecting-logical-fallacies}.

\textit{Output and JSON Structure.} The culmination of these prompts results in a structured JSON output that categorizes identified logical fallacies. This output includes:

\begin{itemize} \item An array of detected logical fallacies, with annotations providing a rationale for each identified instance. \item Sentence ranges indicating where each fallacy occurs within the article, allowing for direct reference. \end{itemize}

\noindent This output structure enables the analysis of the logical coherence of the narrative. Through the application of these specialized prompts, we go beyond fact-checking by supporting reasoning about text, providing a novel solution to the challenges of misinformation in journalism.

\textbf{Conversation Management.} Our framework provides interventions to reduce the potential influence of fallacies on readers (\textbf{UIDO1}), including live chats with the LLM for consulting more information about the detected fallacies. The LLM layer manages the conversation between a reader and the model by handling the chat API and memorizing the conversation history.
\section{Interface Design}
\label{section:interface-design}

\begin{figure*}[t]
  \includegraphics[width=\textwidth]{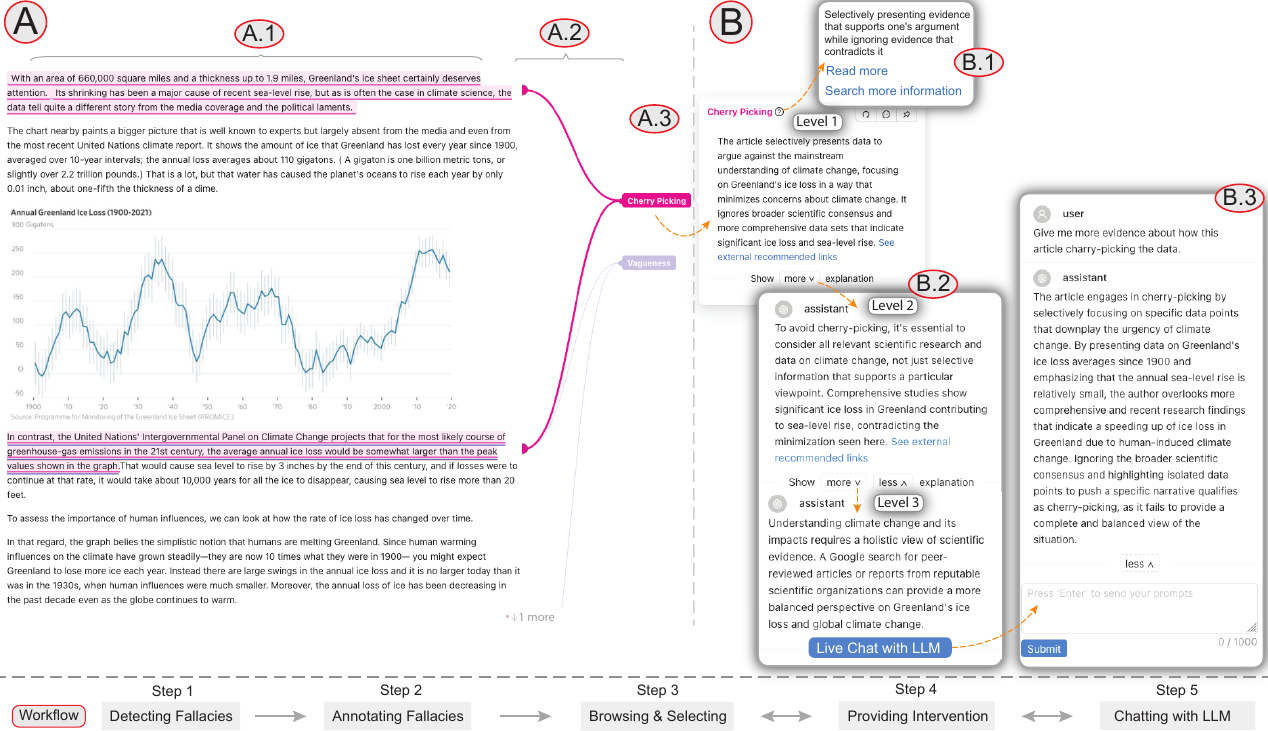}
  \caption{Our two-module interface for detecting and annotating fallacies in news. In the fallacy and text-chart linkage annotation module (A), there are three components: (A.1) inline annotations for the original news content, (A.2) content-fallacy linkage, and (A.3) fallacy tags. The multi-layer interventions (B) in the popup panel for each fallacy has three functional parts: (B.1) the explanation of the selected fallacy, (B.2) the three-layer fallacy correction, and (B.3) live chatting with LLM.}
  \Description{(Interface Figure) The figure shows the interface screenshot and the workflow demonstration of the framework. Part A shows the fallacy and text-chart linkage annotation module. Part B shows the multi-layer interventions in a popup panel for each fallacy. The workflow part shows the interface's operation workflow.}
  \label{fig:teaser}
\end{figure*}

Our framework interface contains two modules: 1) fallacy annotations (\textbf{SDO2}, Figure~\ref{fig:teaser}(A)) and 2) multi-layer interventions to reduce the influence of the detected fallacies (\textbf{UIDO1}, Figure~\ref{fig:teaser}(B)). To enable readers to recognize misinformation more effectively, the core concept of the interface design lies in building dynamic interactive linkages between a set of fallacy-fixing interventions and the text blocks that have been identified as containing potential fallacies. 
\addition{During the development lifecycle, we closely collaborated with a communications expert through an iterative and formative design process. At each stage, we presented early prototypes, functional specifications, and design rationales for individual components. Feedback sessions were conducted regularly to verify the theoretical alignment, practical relevance, and clarity of each proposed feature. This collaboration ensured that the framework's development was grounded in established communication theories and refined based on expert insights into effective misinformation resistance strategies.}

\textbf{Reader's Workflow.}

Our framework supports five operations in the reader's workflow (Figure~\ref{fig:teaser} Workflow). After extracting the news content and detecting potential fallacies (\textbf{step 1}), our interface displays an overview of all the detected fallacies with visual marks on the related text blocks (\textbf{step 2}). The interface allows readers to browse the overall fallacy distribution in the article and filter the fallacy they are interested in by clicking an annotation (\textbf{step 3}). The interface then provides interactive information generated by the LLM as interventions to potentially mitigate the influence of the selected fallacy (\textbf{step 4}). If readers are not satisfied with the LLM-generated information, they can regenerate it or adjust the information through the LLM chat (\textbf{step 5}).

\subsection{\addition{Potential} Fallacy Annotation} \label{subsection:fallacy_annotation}

In this module, we use interactive visual components and a context-preserving design~\cite{6064990} to link each potential fallacy to the corresponding text block in the news article (Figure~\ref{fig:teaser}(A)). To make each fallacy type distinguishable from others, we use an 8-class qualitative color scheme~\cite{10.1179/000870403235002042} that assigns different colors to the \addition{potential} fallacies belonging to different categories, as presented in Figure~\ref{fig:fallacy-table}.

The module has three components: 1) the inline annotation (Figure~\ref{fig:teaser}(A1)) on the original news text and chart content, 2) the content-fallacy linkage (Figure~\ref{fig:teaser}(A2)), and 3) the fallacy tags (Figure~\ref{fig:teaser}(A3)). Each \addition{potential} fallacy detected in the article is connected to its corresponding text block using these components.

\textbf{Inline Annotation.} The \addition{potential} fallacy detection result from the model contains sentence ranges corresponding to the text blocks where each fallacy occurs. We use colored underlines to interactively mark the sentences that are considered to have fallacies (Figure~\ref{fig:teaser}(A1)). When a reader hovers or clicks on a sentence with such inline annotation, the sentence background will be highlighted in the same color as the fallacy linked to it. Meanwhile, the whole linkage from the selected sentence to its fallacy tag will be highlighted synchronously to help the reader focus on the selected content and its associated fallacy.

\textbf{Content-Fallacy Linkage.} Multiple sentences may share the same fallacy type but appear in different paragraphs of the article. To effectively link each fallacy tag and its corresponding interventions to the related text block(s), we establish dynamic visual links using cubic Bézier curves~\cite{4015425} (Figure~\ref{fig:teaser}(A2)). To minimize the occlusion of the selected content-fallacy linkage and make the target linkage stand out from surrounding information, we highlight the clicked linkage by adjusting the line opacity. The color of each link remains the same as the fallacy color. The lines that link the content blocks and the fallacy tags are dynamically updated with the web page scrolling to maintain an intuitive linkage.

\textbf{Fallacy Tags.} The right-side terminal of the content-fallacy linkage is the fallacy tags (Figure~\ref{fig:teaser}(A3)), which display the detected fallacy names. Readers can click on a fallacy tag, and the whole linkage of this fallacy will be highlighted to make the corresponding text block(s) visually salient. A popup panel containing the hierarchical fallacy intervention information can be triggered by clicking.

\subsection{Fallacy Intervention}
\label{subsection:fallacy_intervention}

Guided by the misinformation correction methods from the communication domain~\cite{cook2015misinformation, SwireEcker+2018+195+211, 10.1093/joc/jqz014}, our framework provides a set of interactive intervention approaches in the popup panel for each \addition{potential} fallacy annotation (Figure~\ref{fig:teaser}(B)). Inspired by the dynamic error-explanation linkage design from Lei et al.'s linter system~\cite{10273434}, we support instant explanations and correction information beside each detected fallacy like a linter framework automatically with the help of LLM. The intervention contains three functional parts: 1) an explanation of the selected fallacy (Figure~\ref{fig:teaser}(B1)), 2) a three-layer fallacy correction (Figure~\ref{fig:teaser}(B2))
, and 3) the live chat with \addition{an} LLM in the context of the selected fallacy (Figure~\ref{fig:teaser}(B3)). Readers can regenerate the fallacy explanations and corrections by clicking the ``refresh button'' (\raisebox{0pt}[0pt][0pt]{\raisebox{-0.5ex}{\includegraphics[height=2.3ex]{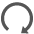}}}).

\textbf{Fallacy Explanation.}
The explanation for a selected fallacy gives readers a better understanding and extra contextual information that facilitates their engagement with the follow-on intervention information. By hovering on the ``question mark'' (\raisebox{0pt}[0pt][0pt]{\raisebox{-0.5ex}{\includegraphics[height=2.3ex]{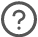}}}) beside the fallacy name, a popup window displays the brief textual explanations. Readers can click "read more..." to check the fallacy definition on Wikipedia. This panel also connects to the search engine API. Readers can search for more information about the fallacy by clicking the "Search more information" link. For a fallacy related to the misinterpretation between the data visualization and text content in an article, we provide an additional explanation for justifying why the fallacy is identified in the selected chart-text linkage.

\textbf{Hierarchical Misinformation Correction Strategy.}
Our approach employs a hierarchical strategy to correct misinformation, inspired by the method outlined by Cook et al.~\cite{cook2015misinformation}. This strategy is divided into three distinct levels of intervention, designed to provide a complete response to detected fallacies.

The first level (Figure~\ref{fig:teaser}(Level 1)), Basic Clarification aims to quickly address the misinformation by providing a straightforward explanation of why the information is misleading. This immediate correction is a necessary first step for correcting basic misconceptions and preventing the spread of misinformation.
The second level (Figure~\ref{fig:teaser}(Level 2)), In-Depth Correction with Evidence, goes beyond clarification by offering additional evidence. This level educates readers and encourages a more critical approach to evaluating information. 
The third and final level, Preemptive Information and Contextual Education (Figure~\ref{fig:teaser}(Level 3)), anticipates common forms of misinformation and addresses them before the user encounters them. A proactive approach enables readers to set up a cognitive guard before this misinformation occurs as it is hard for any information to be retracted (factual or incorrect) and thus, if a preemptive warning is set up, this problem may be mitigated.

To facilitate reader engagement without overwhelming them, our interface allows readers to control the depth of information presented. By selecting ``more'' (\raisebox{0pt}[0pt][0pt]{\raisebox{-0.5ex}{\includegraphics[height=2.3ex]{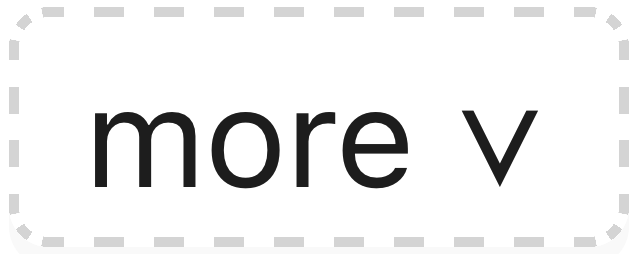}}}) or ``less,'' (\raisebox{0pt}[0pt][0pt]{\raisebox{-0.5ex}{\includegraphics[height=2.3ex]{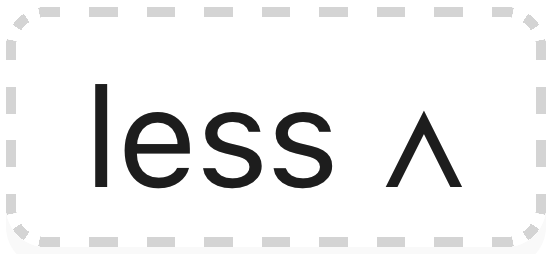}}}) in the interface, readers can navigate through the layers of correction, from basic clarification to in-depth analysis and preemptive education. When applicable, layers contain an outbound link for further exploration on search engines like Google and Bing, supporting lateral reading and cross-referencing of sources for additional information.

\begin{figure*}[t]
\centering	
\includegraphics[width=\linewidth]{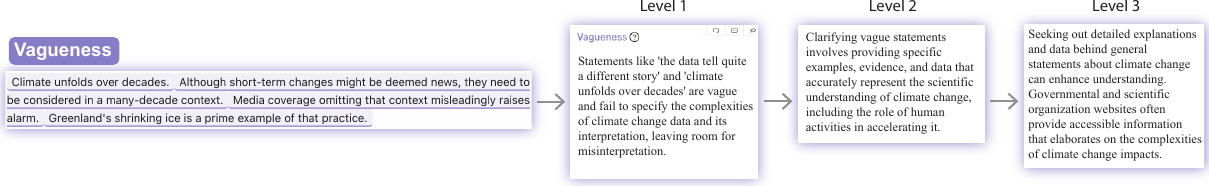}
\caption{Output of case study \#1. The annotations point out a potential Vagueness fallacy in the article, highlighting issues with specificity and the misinterpretation of ice loss trends.}
\Description{(Flowchart) This figure shows the output of the case study in the form of a flow chart. The left side shows the article content with annotations of two detected fallacies. The right side shows the interventions to reduce the influences of these fallacies on readers.}
\label{fig:case1}
\end{figure*}

\textbf{Chatting with the LLM.}
The LLM-generated fallacy explanations may not be able to cover all the necessary parts for a reader to fully understand the detected fallacy issues. There should be a mechanism for readers to flexibly consult with the LLM and get a deeper understanding of the fallacy based on their needs. Our framework provides a live chat function for readers to communicate with the LLM. Readers can type their message and send it to the LLM in the text area at the bottom of the panel (Figure~\ref{fig:teaser}(B3)). The LLM will reply to the reader in real-time right after the reader's message. They can also start chatting by clicking the ``chat icon'' (\raisebox{0pt}[0pt][0pt]{\raisebox{-0.5ex}{\includegraphics[height=2.3ex]{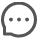}}}) on the panel's toolbar. In order to enhance usability and reduce the learning loads in a reader's fallacy-correction workflow, the content layout for fallacy explanations is similar to the one used in popular chatting apps.

\subsection{Implementation}
Our framework interface is implemented in the form of a web browser extension which provides an immediate and straightforward fallacy detection and correction experience, and meanwhile, maximizes its ability to adapt to a wide range of online news websites. To facilitate the future expansion and maintenance of the framework, we use React\footnote{https://react.dev/} to implement a modularized interface. The interactive content-fallacy linkages and related animations are drawn by D3.js~\cite{bostock2011d3}. 
We use asynchronous APIs to connect to the back-end services \addition{and prevent communication bottlenecks in the workflow}. A Flask-based back-end server\footnote{https://flask.palletsprojects.com} contains the large language model layer functions for communication with the LLM and provides support to multiple LLMs. Our source code is available in this GitHub repository~\footnote{https://github.com/VADERASU/Skeptik-longterm}.

\section{Case Studies}
\label{section:case}

To demonstrate the practical application and effectiveness of our framework, we present three case studies \addition{from the adFontis media dataset that} focused on real-world news stories. We \minor{randomly} selected case studies with moderate bias to show the full extent the capabilities of our system. Upon loading the articles into our framework, the system autonomously detects a wide array of logical fallacies without the need for any manual inputs. \minor{Through the interface, experts explored each annotation interactively, hovering over highlighted text to view suggested fallacies and their definitions. They expand reasoning explanations, review linked sources, and choose whether to accept or dismiss individual annotations, placing them in control of the interpretive process.} \addition{More example articles are provided in Appendix~\ref{appendix:case_appendix}.}

\begin{figure*}[t]
\centering	
\includegraphics[width=\linewidth]{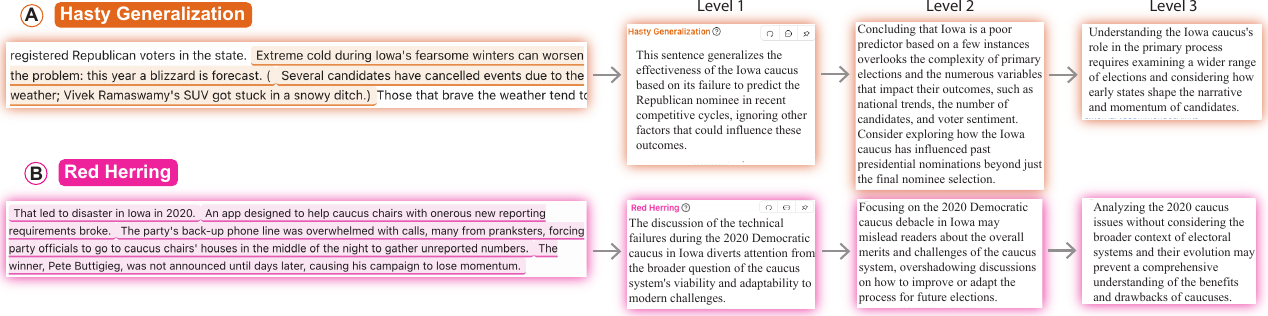}
\caption{
Fallacies present in case study \#2. The Hasty Generalization (A) and Red Herring (B) fallacies are detected and given three annotations, with each level offering more in-depth explanations as to why each excerpt from the article constitutes an occurrence of that fallacy.
}
\Description{(Flowchart) This figure shows the detected fallacies in case study 2. For each fallacy, the figure shows the annotations on the original article content and the three-level intervention.}
\label{fig:case2}
\end{figure*} 

\subsection{Case Study 1: Greenland's Melting Ice Is No Cause for Climate-Change Panic}
\label{subsection:case1}

The article ``Greenland’s Melting Ice Is No Cause for Climate-Change Panic,'' authored by Steven E. Koonin~\cite{koonin2022greenland}, presents an opinion that challenges the prevalent narrative on the impact of climate change on Greenland's ice sheet (Figure~\ref{fig:teaser} and ~\ref{fig:case1}). Published on February 17, 2022, in the Opinion section, Koonin argues against the alarmism surrounding the melting of Greenland’s ice, suggesting that the situation is not as dire as portrayed in the current discourse. He highlights that despite ongoing global warming, the annual loss of ice from Greenland has seen periods of decrease over the past decade, contrary to expectations based on human-induced climate change. 

According to our annotation backend, the piece employs a range of fallacies to support its argument (Figure~\ref{fig:teaser}(A)). Beginning with Cherry-Picking, Koonin selectively presents data to argue against the mainstream understanding of climate change, focusing on Greenland's ice loss in a way that minimizes concerns. This approach ignores broader scientific consensus and data that suggest significant ice loss and sea-level rise. The annotation gives links to external sources that offer evidence contradicting the claims made in the article, advocating for a more balanced and informed perspective on the issue (Figures~\ref{fig:teaser}(B)).

The analysis also addresses \textbf{Vagueness} in Koonin's argumentation. Statements like "Media coverage omitting that context misleadingly raises alarm. Greenland's shrinking ice is a prime example of that practice" lack necessary specificity, making it challenging to grasp the scientific consensus on the impacts of climate change, particularly concerning Greenland's ice melt. This vagueness leaves readers unclear on the specifics of the argument, reducing the article to an unfocused critique.

Furthermore, the article's interpretation of trends regarding Greenland's ice loss over time might be misleading, suggesting that recent decreases in annual ice loss contradict global warming trends without accounting for statistical nuances. Correcting this misconception emphasizes that short-term decreases do not negate the long-term trend of global warming and increased ice loss over time.

This case study highlights the importance of critical reading and recognizing faulty rhetoric when engaging with opinion articles. It underscores the necessity of considering all relevant scientific research and avoiding logical fallacies in argumentation.

\begin{figure*}[t]
\centering	
\includegraphics[width=\linewidth]{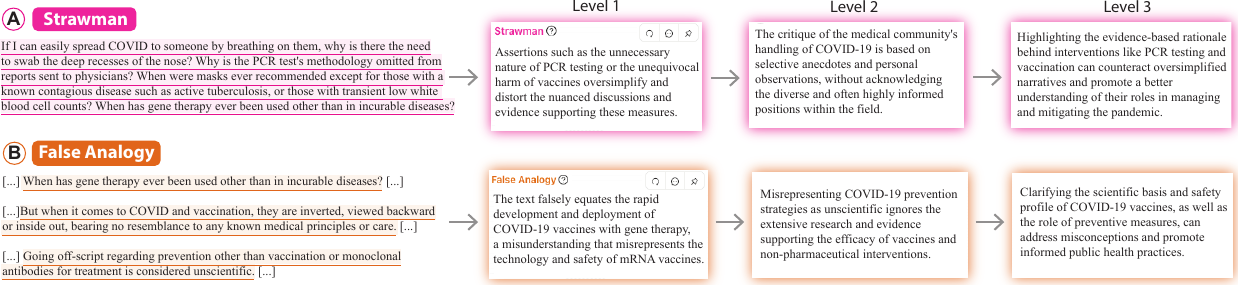}
\caption{
Fallacies present in case study \#3. The Strawman (A) and False Analogy (B) fallacies are detected in the article ``The Medical Profession Implodes.''
}
\Description{(Flowchart) This figure shows the detected fallacies in case study 3. For each fallacy, the figure shows the annotations on the original article content and the three-level intervention.}
\label{fig:case3}
\end{figure*} 

\subsection{Case Study 2: How the Iowa Caucuses Are Changing in 2024}

The article ``How the Iowa Caucuses Are Changing in 2024,'' published in the Economist~\cite{economist2024iowa}, discusses adjustments within the political tradition of caucusing in Iowa, focusing on the shift by the Democrats from an in-person to a mail-in ballot system (Figure~\ref{fig:case2}). (A caucus is  a local, in-person meeting where party members discuss and vote for candidates.) This marks a significant departure from a tradition that dates back to the 1970s, highlighting evolving strategies within American political primaries. The Republicans, however, will follow traditional caucusing in only six states, underscoring a divergence in approach between the two major political parties. The analysis identifies several logical fallacies, including Hasty Generalization, Red Herring, and Vagueness.

First, the Hasty Generalization fallacy (Figure~\ref{fig:case2}(A)) is evident in the article's critique of caucuses based on their cost and low attendance, without fully considering the benefits they offer in terms of voter participation. The excerpt from the original article, ``Extreme cold during Iowa’s fearsome winters can worsen the problem: this year a blizzard is forecast,'' exemplifies how a single factor is exaggerated to undermine the caucus system without acknowledging its broader value to Iowa's political culture. The critique of caucuses for being costly and poorly attended, as highlighted by an example of extreme weather, fails to account for the full picture of what caucuses contribute, thereby making this a Hasty Generalization.

The Red Herring fallacy (Figure~\ref{fig:case2}(B)) in the article diverts attention from the core issues surrounding the future and effectiveness of the caucus system by overly focusing on the technical hardships of the 2020 caucus. Specifically, the article writes about the malfunctioning app and the overwhelmed backup phone system, which led to significant delays in announcing the Democratic caucus results. By concentrating on this isolated incident, the narrative shifts in a way that contributes to the perception of caucuses as disorganized, even though these technical issues may not be directly relevant to the overall effectiveness of the caucus process itself. This diversion acts as a Red Herring, misleadingly suggesting that one technical difficulty is to blame for the caucus system's flaws.

This case study exemplifies the utility of critical analysis in identifying fallacies and responsibly interpreting arguments. It emphasizes the need to consider multiple perspectives and avoid oversimplifications when discussing complex political processes.

\subsection{Case Study 3: The Medical Profession Implodes}

The article "The Medical Profession Implodes," authored by Steve Karp, M.D., and published on December 13, 2021, in the American Thinker~\cite{Karp2021Implosion}, offers a critical perspective on the medical profession's response to the COVID-19 pandemic (Figure~\ref{fig:case3}). Karp suggests that the medical field has abandoned scientific principles in favor of adhering to government mandates and public health policies, which he views as overreaching and not grounded in sound science. He criticizes the rapid adoption of COVID-19 vaccines, likening their development to gene therapy, and questions the efficacy and safety of pandemic-related health measures such as PCR testing, masking, and social distancing.
Our analysis identifies several logical fallacies within Karp's argument, notably Strawman and False Analogy, which undermine the credibility of his critique.

Strawman is evident in Karp's portrayal of the medical community's adherence to COVID-19 guidelines as complete mind control by the government (Figure~\ref{fig:case3}(A)). This misrepresents and oversimplifies the debates within the medical and scientific communities about the best approaches to managing the pandemic. By attacking this oversimplified argument, Karp fails to engage with the actual reasons and evidence that have guided public health recommendations and policies, thus making this argument a strawman.

The False Analogy fallacy is also present in Karp's comparison of the development and deployment of COVID-19 vaccines to gene therapy (Figure~\ref{fig:case3}(B)). This analogy misleads by equating two distinct scientific processes and therapeutic approaches, causing confusion about the nature of mRNA vaccines. Such a comparison ignores the clinical research that mRNA vaccines underwent before their authorization for emergency use. This oversight diminishes the article's argument by conflating unrelated concepts to question the validity of COVID-19 vaccines.

\section{Evaluation}
\label{section:evaluation}

\begin{figure}[t]
\centering	
\includegraphics[width=0.6\linewidth]{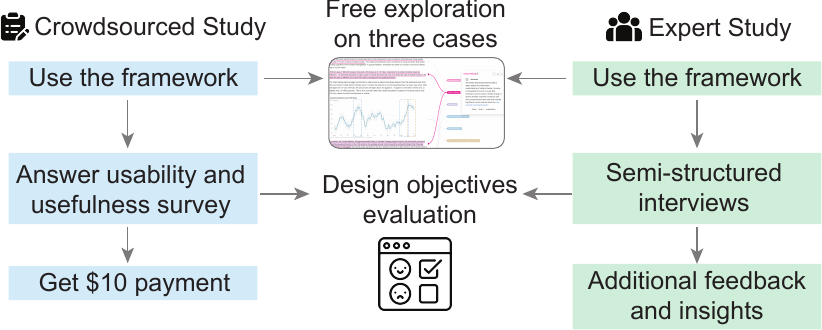}
\caption{Crowdsourced study and expert study pipeline. These refer to evaluations conducted in Sections~\ref{subsection:crowdsourced_study} and ~\ref{subsection:expert_interviews}.} 
\Description{(Flowchart) This figure shows the pipeline of our crowdsourced study and the expert study. The three parts of the crowdsourced study are using the framework, answering usability and usefulness survey, and getting the payment. For the expert study, the three parts are using the framework, the semi-structured interviews, and additional feedback and insights.}
\label{fig:study-pipeline}
\end{figure}

We conducted a non-human quantitative study to assess our prompt's validity. We followed up our evaluation with two separate human evaluation studies (see Figure~\ref{fig:study-pipeline} for the study sequence). The initial phase involves crowdsourcing, focusing on gathering feedback pertinent to our design objectives. 
\addition{The} second part consists of expert interviews to obtain more detailed insights as well as future directions of research. Both parts of the study were approved by the \addition{Arizona State University} IRB Board.

\subsection{Ad Fontes Dataset Correlation Study}
\label{subsection:ad_fontes_study}

One of the ways we decided to evaluate our framework is by using an existing dataset. However, no dataset perfectly suits our needs—for instance, there is no ground truth dataset that explicitly states which sentences in an article contain a specific fallacy, such as a strawman or post hoc fallacy. As a substitute, we analyzed an existing dataset that rates news reliability and bias and compared it to the outputs of our tool. Our reasoning is that \addition{bias in news articles often manifests through the use of fallacious reasoning and misleading data interpretation or visualization. Prior research has shown that practices such as cherry-picking and false balance are common mechanisms through which media bias distorts the audience's perception~\cite{mohseni2022best}. Therefore, we expect that sites rated as more biased will exhibit a higher prevalence of such features.}
\addition{Moreover, since our approach leverages LLMs to detect logical fallacies, this analysis provides a valuable proxy to demonstrate the promise of LLMs in identifying fallacious reasoning. Although our primary goal is to present a novel user interface that facilitates critical thinking~\cite{ecker2022psychological} and close reading~\cite{https://doi.org/10.1002/bin.1512}, we acknowledge the absence of direct comparative benchmarks for LLM-based fallacy detection. Thus, we selected a large-scale dataset with established bias and reliability metrics to indirectly validate the effectiveness of our method.} This analysis represents the best option for a large-scale quantitative study using expert coders to evaluate the data.

\begin{figure}[t]
\centering	
\includegraphics[width=\linewidth]{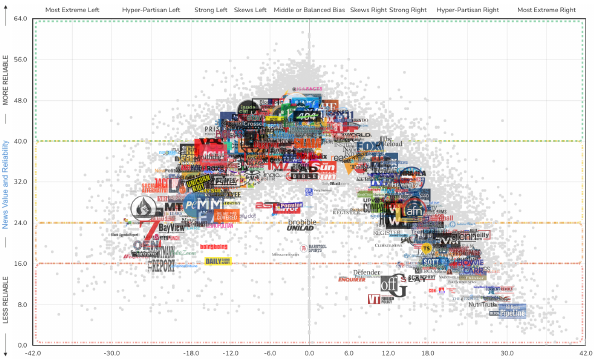}
\caption{The Ad Fontes Media Bias Chart. News Sources are plotted on an x-y axis, with x-axis values denoting left or right-leaning sources. y-axis values denote the reliability or overall credibility.}
\Description{a screenshot of the Ad Fontes Media Bias Chart contains the distribution of news sources regarding their reliability and bias scores.}
\label{fig:adfonte}
\end{figure}

We purchased data from Ad Fontes Media~\footnote{https://adfontesmedia.com/interactive-media-bias-chart/}, which includes bias and reliability scores for over 80,000 online articles. Negative bias values correspond to left-leaning articles, with -30.0 representing the most extreme left-wing perspectives. Positive bias values indicate right-leaning articles, with +35.0 representing the most extreme right-wing perspectives. Higher reliability scores indicate greater reliability, with values ranging from 8 to 48 across the articles. There is a notable correlation between reliability and bias: as bias becomes more extreme (either positive or negative), reliability tends to decrease. Figure~\ref{fig:adfonte} provides an overview of the Ad Fontes Media Bias Chart. The bias and reliability scores are rated by trained analysts with a standardized methodology.

\begin{figure*}[t]
\centering	
\includegraphics[width=0.8\linewidth]{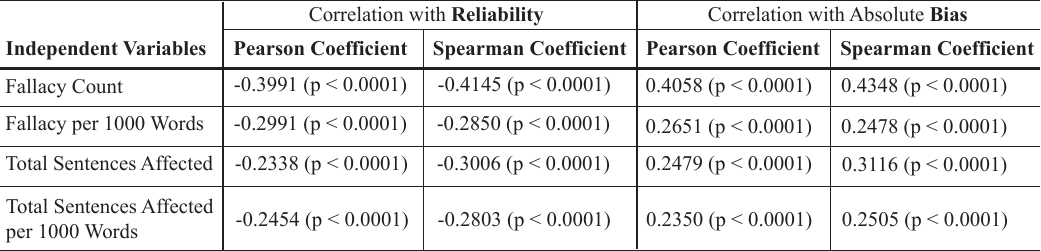}
\caption{
 Pearson and Spearman coefficients were reported for numerical independent variables. The results show a significant negative correlation between reliability and fallacy count, indicating that as reliability increases, the number of fallacies decreases. Additionally, there is a positive correlation between the absolute value of bias and the fallacy count, suggesting that greater bias is associated with a higher number of fallacies. This is expected, as there is an innate correlation within the dataset indicating that the absolute value of the bias and reliability are negatively correlated.}
\Description{(Table) This table shows the Pearson coefficients and Spearman coefficients between the four independent variables and the dependent variables which are the reliability and absolute bias.}
\label{fig:correlations}
\end{figure*} 

\noindent \textbf{Experiment Setup:} Our study aimed to validate whether the logical fallacies predicted by our GPT model are correlated with the reliability and bias of the news articles in the dataset. To achieve this, we conducted a set of quantitative experiments, including correlation analysis, fallacy distribution analysis, and regression modeling. We used the following parameters in our experiments. For dependent variables, we used the following

\begin{itemize}
    \item \textit{reliability}: A numeric value ranging from 8 to 48, with higher scores indicating greater reliability.
    \item \textit{bias}: A numeric value, with negative values for left-leaning bias and positive values for right-leaning bias. For analysis, we used the absolute value of bias.
\end{itemize}

\noindent For the independent variables, we used the following:

\begin{itemize}
    \item \textit{fallacy\_count}: The total number of logical fallacies detected in an article (numeric, $\geq$0).
    \item \textit{fallacies\_per\_1000\_words}: The number of logical fallacies detected per 1000 words in an article (numeric, $\geq$0).
    \item \textit{total\_sentences\_affected}: The total number of sentences in an article that contain logical fallacies (numeric, $\geq$0).
    \item \textit{total\_sentences\_affected\_per\_1000\_words}: The number of sentences containing fallacies per 1000 words (numeric, $\geq$0).
    \item For each of the following fallacies, we used a binary indicator (0 or 1) representing the presence of the fallacy in an article:
    
    \textit{EBP} (Evading the Burden of Proof), \textit{ST} (Strawman), \textit{RH} (Red Herring), \textit{CP} (Cherry Picking), \textit{FA} (False Analogy), \textit{HG} (Hasty Generalization), \textit{PH} (Post Hoc), \textit{FC} (False Cause), \textit{VAG} (Vagueness).
\end{itemize}

\noindent \textbf{Hypotheses:}
Our hypotheses are twofold:
\begin{itemize}
    \item \textbf{H1}: There will be a negative correlation between the reliability and the number of fallacies per 1,000 words. This is because we anticipate that as the reliability of the news articles increases, the fallacies will decrease.
    \item \textbf{H2}: There will be a positive correlation between the absolute value of the bias and the number of fallacies per 1,000 words. This is because as the news article is more polarized, the number of fallacies present in the article will also increase. This also follows from H1, since there is a negative correlation between the reliability and the absolute value of the bias. See Figure~\ref{fig:adfonte} (Ad Fontes Figure).
\end{itemize}

We analyzed a random sample of 3,825 articles from a dataset of 80,000 articles. The sample was selected to contain bias and reliability values that span the full spectrum. The significance level for our statistical analyses was set at $\alpha=0.05$, corresponding to a 95\% confidence level.

We performed Pearson and Spearman correlation analyses to assess the relationships between the fallacy-related features and the reliability and bias scores of the articles.

Fallacy count, fallacies per 1,000 words, total sentences affected, and total sentences affected per 1,000 words all showed a negative Pearson and Spearman correlation with reliability (see Figure~\ref{fig:correlations}), indicating a strong negative relationship between the number of fallacies and article reliability. More reliable articles contain fewer fallacies, and the number of fallacies, when normalized for word count, also decreases as reliability increases. Specifically, the correlation between fallacies per 1,000 words and reliability is -0.2991 for Pearson and -0.2850 for Spearman.

Additionally, the correlation between the absolute value of the bias and fallacies per 1,000 words is 0.2651 for Pearson and 0.2478 for Spearman, suggesting that as bias increases, the presence of fallacies also tends to increase. \addition{This finding supports our assumption that biased content frequently employs flawed reasoning, making it particularly suitable for evaluating the efficacy of our fallacy detection approach.}

\begin{figure*}[t]
\centering	
\includegraphics[width=0.8\linewidth]{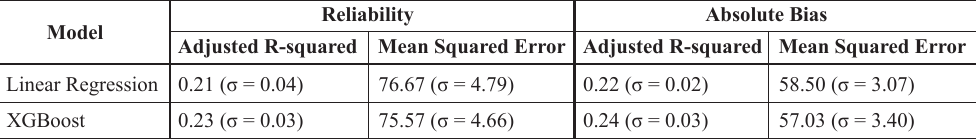}
\caption{
The average adjusted R-squared and mean squared error values with standard deviations for the 5-fold cross-validation for linear regression and XGBoost models. The adjusted $R^2$ values are relatively low (around 0.21 to 0.24), which means that the models explain only a small portion of the variance. 
}
\Description{(Table) This table shows the adjusted R-squared and the mean squared error values with standard deviations for the 5-fold cross-validation of the linear regression and XGBoost models. The dependent variables are the reliability and the absolute bias.}
\label{fig:model_results}
\end{figure*}

Finally, we applied both linear regression and XGBoost models to explore the relationship between fallacy-related features and the reliability and bias scores. We found that for bias, $MSE=58.50$, adjusted $R^2=0.24$ and for reliability, $MSE=76.67$, adjusted $R^2=0.23$ under the linear regression model. For the XGBoost model, $MSE=57.66$, which indicates a marginal decrease in error when compared to the linear regression models. These models (see Figure~\ref{fig:model_results}) confirm the significant impact of logical fallacies on the reliability and bias. We note that the adjusted $R^2$ values are relatively low, being around 0.21-0.24. Due to the relatively few features that our model employed (as listed in \textbf{Experiment Setup} in Section~\ref{subsection:ad_fontes_study}), this is not abnormal, and a future large scale study could explore using actual article text, rhetorical devices and writing style to better predict bias and reliability. The MSE values range from 57.03 to 76.67, with XGBoost having slightly lower MSEs compared to the linear regression models, meaning that XGBoost was better at minimizing the prediction errors. Additionally, the standard deviations of the MSEs are relatively low compared to the MSE values, indicating that there is a consistent performance across the cross-validation folds.

In summary, both \textbf{H1} and \textbf{H2} are supported, as there is a negative correlation between the reliability and the number of fallacies per 1,000 words. There is also a positive correlation between the absolute value of the bias and the number of fallacies per 1,000 words (Figure~\ref{fig:correlations}. Both of these were confirmed with linear regression and XGBoost models (Figure~\ref{fig:model_results}).

\subsection{Crowdsourced Survey}
\label{subsection:crowdsourced_study}

Participants for our study were recruited via Prolific.co, with eligibility criteria including U.S. residency, desktop computer usage,
and a minimum 95\% task acceptance rate on Prolific.co. They were instructed to engage with the system through a structured process that included an introductory overview, interaction with a demo via case studies, and completion of a usability survey. 50 participants were recruited. They took an an average of 23 minutes 48 seconds to read through the case studies and answer our survey. We paid them \$10 upon successful completion of the study.

The survey's questions, inspired by the Unified Theory of Acceptance and Use of Technology (UTAUT)~\cite{venkatesh2008technology}, aimed to evaluate the system's efficacy in enhancing article comprehension, detecting misinformation, and identifying logical fallacies. These questions were categorized into Performance, User Interface (UI), Language Learning Models (LLM), and Usability. The survey used a Likert scale from 1-5, indicating strongly disagree to strongly agree, with higher values indicating better usability and performance.

\begin{itemize}
    \item \textbf{Performance:} This category evaluates the system's influence on users' article understanding, misinformation detection capabilities, detection speed, reading effectiveness, credibility assessments, and grasp of logical fallacies. This aligns with \textbf{SDO1: Identifying potential fallacies in online news automatically.}
    \item \textbf{UI:} Questions in this section assess the utility of visual aids—like sentence underlining, dynamic links between article contents and fallacy tags, background color alterations, and pop-up explanations—in identifying fallacies. This aligns with \textbf{SDO2: Annotating detected fallacies (and other relevant information) dynamically.}
    \item \textbf{LLM:} Here, the focus is on user experiences with language learning models (e.g., GPT), specifically the trustworthiness of fallacy detection outcomes and explanations, along with the effects of real-time interactions.
    \item \textbf{Usability:} 
    This segment examines the ease-of-use of the framework, learning curve, clarity of interactions, annotation recognizability, information readability in pop-ups, ease of accessing desired information through GPT interactions, and overall contentment with the system's integration into the article reading workflow. Usability aligns with \textbf{SDO2} and \textbf{UIDO1: Providing interventions to effectively reduce the fallacy influence.}
\end{itemize}

\begin{figure}[t]
\centering	
\includegraphics[width=0.9\linewidth]{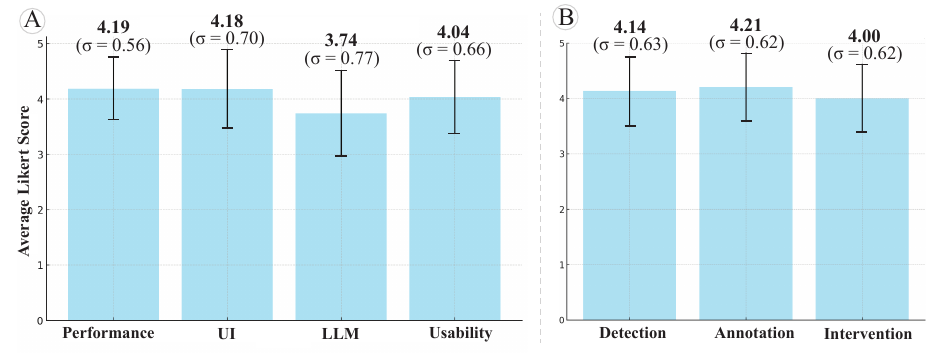}
\caption{\addition{Average survey scores with corresponding standard deviations. (A) presents the scores based on four primary evaluative dimensions: Performance, UI, LLM, and Usability. (B) presents the evaluation scores aligned with the three key phases of our workflow: Detection, Annotation, and Intervention.}}
\Description{(Bar Chart with Standard Deviation Intervals) This figure shows the bar chart of the average survey scores on each evaluate dimension.}
\label{fig:study-result}
\end{figure}

Overall, the quantitative analysis, Figure~\ref{fig:study-result}(A), reveals a preference for the system's UI and usability components, with mean scores surpassing 4. The LLM category, while still positively received, yielded a marginally lower average score of about 3.74. Standard deviations across all categories show moderate response variability. 

\addition{To align more closely with the three major phases in the workflow—Fallacy Detection, Annotation, and Intervention—we also analyzed the survey results according to this conceptual framework. Figure~\ref{fig:study-result}(B) presents this breakdown. The \textbf{Fallacy Detection \minor{phase} (SDO1)}, which encompasses system capabilities for identifying logical fallacies (e.g., performance in fact-checking and LLM outputs), achieved a mean score of 4.14 with a standard deviation of 0.63, indicating a satisfied detection performance. The \textbf{Fallacy Annotation \minor{phase} (SDO2)}, which includes tasks like highlighting and labeling of fallacious content, yielded the highest average score of 4.21 with a standard deviation of 0.62, suggesting that participants particularly valued the framework’s clarity and support in interpreting the content. The \textbf{intervention phase (UIDO1)}, which evaluates both the quality of the generated intervention content and users’ ease of accessing it, received a mean score of 4.00 with a standard deviation of 0.62, reflecting generally positive, though slightly more varied, participant responses.} The full survey measures and results are shown in \minor{Appendix}~\ref{appendix:study_results}.

\subsection{Expert Interviews}
\label{subsection:expert_interviews}

Our methodology also incorporated feedback from a panel of experts (\textbf{E1-E5}) to assess the efficacy of our tool. \textbf{E1}, with expertise in computing and misinformation studies, alongside \textbf{E2-E5}, subject matter experts in communications, provided feedback into the tool's design and functionality.
We conducted semi-structured interviews, allowing each expert approximately 30 minutes to engage with the tool and examine three case studies presented in Section~\ref{section:case}. This was followed by a series of questions aligned with the design objectives outlined in Section~\ref{subsection:design_obj}. The interviews were recorded, and transcriptions were later analyzed. Some experts also submitted written feedback as well. We detail the assessment by each expert of our tool's alignment with the design objectives, acknowledging its imperfections. For every objective, we discuss accompanying qualified concerns or limitations. These limitations and potential future work are further discussed in Section~\ref{section:discussion}.


\textbf{SDO1: Identifying potential fallacies in online news automatically.}
Feedback on the tool's ability to automatically identify fallacies was predominantly positive. For instance, \textbf{E2} highlighted its effectiveness in detecting subtle fallacies, stating, ``It caught some things that I don't necessarily think I would have caught,'' and affirming, ``it helped me identify them more than I probably would have.'' Similarly, \textbf{E3} appreciated its specificity in identifying logical fallacies, which helps in pinpointing reasoning flaws: ``I like that it identifies logical fallacies specifically it helps to identify, you know, the problem in the reasoning rather than just arguing that the facts of the matter.'' However, \textbf{E4} raised concerns about the tool's domain-specific effectiveness:``Just because I might be an so-called expert on misinformation, I mostly study Chinese propaganda. I don’t know much about misinformation surrounding the Iowa caucus or climate change.'' This feedback suggests that while the tool is effective in identifying fallacies, its performance may vary across different subjects, especially without the input of someone with specialized knowledge in those areas.

\textbf{SDO2: Annotating detected fallacies (and other relevant information) dynamically.} 
The dynamic annotation feature received support for its potential to encourage readers to examine articles critically. \textbf{E3} commended the tool for visual indicators of fallacies: ``I like how it has the, you know, the underlining, and then you can top up this way. There is you can see the flags kind of fallacies that are like an article.'' \textbf{E4} praised its capacity to alert readers to misleading elements: ``Your tool catches them and can give reminders to the reader like a `hey watch out for this little trick'.'' Both of these comments support our objective of annotating the fallacies along with helpful information.

\textbf{UIDO1: Providing interventions to effectively reduce the fallacy influence.} Experts noted the tool's effective use of context and counterarguments to mitigate fallacy influence. \textbf{E2} provided feedback on the tool's strategy for delivering context and counterargument: ``it really, in my mind, did a good job of... discrediting the arguments.'' Additionally, \textbf{E2} praised the feature that encourages exploration of outbound links for further information, remarking, "those outbound links... were meant to be similar to, you know, someone opening up a... tab for, you know, alternate
sources and information." This feature aligns with our objective to promote lateral reading practices, which diminishes the persuasiveness of misinformation. Conversely, \textbf{E4} raised a caution about the potential for backlash from fact-checking efforts, noting, ``Fact-checking can also lead to backlash. There’s a lot of literature on this.'' This comment brings to light concerns regarding the reception of credibility systems by the public. \textbf{E2} also commented on this concern, and this topic will be further discussed in Section~\ref{section:discussion}.

\textbf{SDO3: Adapting to varying fallacy types.} 
The primary focus of this design objective is detailed in Section~\ref{subsection:classification_for_logical_fallacies}. Nonetheless, \textbf{E1} offered insights beyond merely adapting to various fallacy types, suggesting the tool's flexibility for integrating different components or utilizing alternative frameworks. \textbf{E1} noted, "And this system is pretty flexible, which means we can connect to any, any, you know, large language models through the API." Furthermore, they added, "we can, you know, set different parameters and temperatures for the job and say like, if you change the temperature, you should get some consistency in there." These remarks, although not explicitly addressing the tool's adaptability to varying fallacies, suggest an underlying versatility in the design that paves the way for future enhancements and integrations.

\section{Discussion}
\label{section:discussion}

In this section, we compile insights and feedback from our evaluations and expert interviews, which confirm the effectiveness of our framework in a real-world context. \addition{We begin by discussing the use of LLMs in logical fallacy detection, highlighting both their potential and inherent limitations. Several additional limitations and suggestions for future work also emerged from the expert interviews; we incorporate these perspectives to provide a cohesive and comprehensive discussion. Finally, we outline directions for future research.} 

\subsection{\addition{Limitations Regarding LLM Reliability and Fallacy Detection}}

\addition{While LLMs have demonstrated remarkable capabilities, their integration into our framework also introduces notable limitations that we acknowledge.}

\textbf{Overalignment:} As noted by \textbf{E1}, some LLM-generated annotations were perceived as “vague” and insufficiently helpful. This reflects a broader tendency of LLMs toward overalignment~\cite{sun2024trustllm}, where models prioritize caution to minimize the risk of producing offensive, incorrect, or misleading content. Overalignment often results in outputs that are technically safe but lack specificity or actionable insight, particularly when faced with nuanced queries.

\textbf{Risk of Hallucinations:}
\addition{
A further limitation stems from the potential for hallucinations—where LLMs generate outputs that are factually incorrect or misleading, especially when interpreting complex or context-sensitive information~\cite{10.1145/3703155}. To mitigate this risk, we carefully framed the tool's outputs as suggestions rather than definitive conclusions. Annotations are phrased tentatively (e.g., “This may be an example of...”) to encourage users to critically evaluate the system’s feedback rather than accept it uncritically. This approach acknowledges the inherent possibility of both false positives and false negatives in LLM-assisted fallacy detection. \minor{We acknowledge that this mechanism alone is not a sufficient safety net. Relying solely on users to discern incorrect or incomplete annotations may compromise the robustness and reliability of the system in high-stakes or domain-specific scenarios.}
}

\textbf{\minor{Limited Domain Generalization:}} 
\minor{As raised by Expert \textbf{E4}, the effectiveness of \textit{Skeptik} may vary across topics such as climate science, healthcare, or politics. In domains that require specialized knowledge, general-purpose LLMs may fail to accurately detect nuanced fallacies or provide sufficiently context-aware explanations. Without access to domain-specific corpora or integration with retrieval-augmented generation (RAG) pipelines, the system may misinterpret key terminology or overlook context-specific reasoning patterns.}

\textbf{\addition{Evaluation Scope:}}
\addition{Our primary objective is to explore how LLMs can assist users in identifying logical fallacies through an interactive interface, rather than to benchmark LLM accuracy in fallacy detection—a task hindered by the absence of standardized datasets. To provide preliminary validation, we conducted a study (Section 6.1) analyzing media sources with known biases and trustworthiness levels. Findings indicate that sources with higher bias and lower credibility exhibited more detected fallacies, suggesting the potential utility of LLMs in this domain.}

\addition{By integrating LLMs into a human-in-the-loop framework, we aim to support users in developing critical thinking skills and engaging in close reading practices. Overall, while a rigorous evaluation of LLM performance in fallacy detection would be an important contribution—likely best addressed in a dedicated benchmark paper—the focus of this work is to demonstrate how LLMs can be harnessed to assist human judgment. We anticipate that as LLM capabilities continue to advance, frameworks like ours will become increasingly valuable. Our modularized design is flexible to adapt to the varying prompts and LLMs. Consequently, our rigorous evaluation centered on the effectiveness of the user interface and the workflow of the fallacy detection, annotation, and intervention, as detailed in our quantitative and qualitative study.}

\subsection{Broader Limitations and Challenges}
\label{subsection:limitations}

\textbf{Engagement and Backfire Concerns:}
While our tool aims to educate users and reduce the influence, there is a noted potential for reader backlash. \textbf{E4} warned of the possibility that our interventions may annoy some readers but could also deepen their skepticism or lead to dismissal of the tool. Earlier research identified the backfire effect, noting that in some cases corrections strengthen rather than weaken beliefs in incorrect information~\cite{nyhan2010corrections}. More recent research has shown that corrections are at least somewhat effective at increasing belief accuracy~\cite{nyhan2021backfire}. Thus, in order for this tool to be effective, there must be a balance between providing corrective information without alienating the audience. Along this vein, \textbf{E2} points out a study by Elkins et al.~\cite{elkins2013users} that investigates how experts respond to credibility assessment systems that challenge their expertise, particularly in the context of deception detection. By applying self-affirmation theory, the study explores whether affirming an individual's self-worth in areas unrelated to their professional expertise can reduce defensiveness and improve openness to the system's recommendations. The findings reveal that while self-affirmation did not increase the accuracy of deception detection, it did enhance experts' objectivity towards and perception of the system, particularly when it offered contrary-minded recommendations. The findings suggest that psychological threats from expert systems can affect user interaction and acceptance, but strategies like affirmation might mitigate these effects and enhance system perceptions. This is a major consideration that needs to be taken into account when designing future iterations of our system. 


\textbf{Narrow Focus on Logical Fallacies:} \textit{Skeptik} primarily targets logical fallacies within textual content and between text and charts. This focus does not encompass other forms of misinformation, such as image manipulation, deepfakes, or coordinated disinformation campaigns. Consequently, the tool may not detect all types of misleading information present in news articles.

\textbf{\addition{Cross-Lingual Applicability:}} 
\addition{\textit{Skeptik}'s current implementation has been evaluated exclusively on English-language content. While the framework's architecture is designed for adaptability, extending its application to low-resource languages presents notable challenges. These include the scarcity of annotated datasets and potential variations in argumentation structures across different cultures and languages. Such factors may affect the accurate detection, classification, and explanation of logical fallacies. Future work should explore cross-lingual transfer learning techniques and culturally informed adaptations to enhance \textit{Skeptik}'s effectiveness in diverse linguistic contexts.}

\textbf{Evaluation Constraints:} Our expert evaluation involved only five participants, which may not provide a comprehensive view of the tool's effectiveness across diverse expert opinions. Additionally, the quantitative analysis, while indicating correlations, does not establish causation or account for all variables influencing article reliability and bias. Also, while we have established a correlation between the frequency of fallacies and the reliability and bias of articles, we have not proven that the fallacies detected are indeed present in the articles. Given the satisfactory reasoning ability of GPT and the examples provided, we are fairly confident that the detected fallacies correspond to actual fallacies in the articles. However, confirmation from domain experts would provide stronger validation.

\subsection{Future Work}

\textbf{Rigorous Evaluation and Longitudinal Studies}: As mentioned above in \textbf{Evaluation Constraints}, our evaluation was limited in scope, involving a small number of experts and focusing on short-term assessments. Future studies should involve a larger and more diverse group of experts and users to validate the tool's effectiveness across different demographics and contexts. \minor{Future work also includes exploring how \textit{Skeptik} can be integrated into the workflows of experts who currently use other misinformation detection methods, in order to better identify integration pathways with existing expert toolchains.}
Conducting longitudinal studies would allow us to assess the impact of sustained tool usage on media literacy and critical thinking skills over time. This comprehensive evaluation would help determine the long-term benefits and potential areas for improvement in \textit{Skeptik}. \minor{Furthermore, we recognize the importance of systematically comparing \textit{Skeptik} to existing misinformation detection methods, including discriminative models (e.g., binary classifiers trained explicitly for misinformation detection) and generative models (LLM-based systems). A benchmark study, similar to the comparative evaluation of discriminative and generative models~\cite{raza2025fake}, would help clarify the strengths and limitations of each approach and further establish the effectiveness of human-in-the-loop generative models in logical fallacy detection contexts.}

\textbf{Domain Adaptability:} Based on feedback from \textbf{E1}, future research will focus on enhancing and \minor{evaluating} our tool's domain adaptability specifically for the detection of logical fallacies \minor{across a wider range of real-world contexts}. This can be achieved by fine-tuning our model using few-shot learning with annotated examples of less common or domain-specific fallacies. We also plan to refine the model's overall performance by optimizing the API and adjusting key parameters such as the LLM temperature. This will help manage the variance in model outputs, thus improving the accuracy and relevance of the fallacy annotations. \minor{In addition, we plan to test \textit{Skeptik} on more diverse content domains, including health, politics, and finance, by curating domain-specific article sets and measuring both annotation quality and user perception across topics with varying complexity. These evaluations will help us better understand how well the system generalizes and where targeted improvements are needed.}
Expanding the classification of fallacies is another straightforward future direction. This involves incorporating a wider range of fallacies into the training data, which will enable our tool to identify fallacies not covered in the current version. 

\textbf{Enhancing Fallacy Detection and Intervention with External Context:} 
Some fallacies (e.g., Cherry Picking, False Cause, Hasty Generalization) are inherently context-dependent and would significantly benefit from external context retrieval. Integrating Retrieval-Augmented Generation (RAG) or similar information retrieval techniques could notably enhance \textit{Skeptik}’s detection and intervention capabilities for these context-sensitive fallacies. Future iterations of our system could incorporate such external information retrieval approaches, aligning with \textit{Skeptik}’s modular and scalable design philosophy and further improving its practical effectiveness.

\textbf{Personalized Approach with Cognitive Biases:} \textbf{E5} suggested that a personalized approach to mitigating misinformation is helpful, pointing out the cognitive biases codex~\cite{manoogian2020cognitive}, which categorizes 188 cognitive biases as a foundation for its detection framework. These biases each represent a pattern of deviation from rationality in judgement, and vary from each person to person. The study on the cognitive biases codex underscores the importance of cognitive bias detection in addressing misinformation and enhancing decision-making processes. The authors suggest that the ability to automatically detect and measure cognitive biases could significantly advance our understanding of human cognition and improve information quality across digital platforms, and thus it can offer a more targeted and effective way to help readers pinpoint and correct misinformation. 
\addition{In addition, in our future work, we plan to explore the integration of a cognitive biases checklist proposed by Draws et al.~\cite{Draws_Rieger_Inel_Gadiraju_Tintarev_2021} into our evaluation workflow to refine the experimental setup and report on potential biases in the studies.}


\textbf{\addition{Assessment with UTAUT:}}
\addition{While the NASA Task Load Index (NASA-TLX) has been more commonly used in visual analytics (VA) research to assess cognitive workload~\cite{hart2006nasa}, it does not directly address factors influencing technology acceptance and usage behavior. In contrast, UTAUT provides a comprehensive model of user acceptance, encompassing performance expectancy, effort expectancy, social influence, and facilitating conditions, yet its application in VA studies remains limited. By incorporating UTAUT into our evaluation, we seek to capture a broader range of user experiences, including cognitive workload, behavioral intentions, and acceptance factors. We encourage future VA research to employ UTAUT alongside traditional measures such as NASA-TLX for a more holistic understanding of user interaction and system adoption.}
\section{Conclusion}
\label{section:conclusion}

In this paper, we present a novel framework aimed at enhancing the credibility of journalism by mitigating potential misinformation through the integration of LLMs and heuristic analysis. By developing a web browser extension that annotates and analyzes online news articles for logical fallacies, we offer a comprehensive tool that aids readers in critically evaluating news content. Our research, supported by case studies, a crowdsourced study, and expert interviews, demonstrates the framework's effectiveness in identifying and annotating misleading information, thereby contributing to the reduction of misinformation spread.


\bibliographystyle{ACM-Reference-Format}
\bibliography{main}

\appendix
\section{LLM Prompts}
\label{appendix:llm-prompts}

\subsection{Prompt for Detecting Logical Fallacies}
\label{appendix:prompt-for-detecting-logical-fallacies}

You are an expert skilled in detecting logical fallacies and explaining why these fallacies occur. Here are 9 fallacies.

Evading the Burden of Proof (EBP): This fallacy occurs when someone makes a claim but refuses to provide evidence to support it, shifting the burden of proof to others.
Example: A politician claims that a new policy will improve the economy but does not provide any data or reasoning to support this claim.

Strawman (ST): Misrepresenting someone's argument to make it easier to refute than the original argument.
Example: Person A says we should have stricter gun control. Person B responds by saying Person A wants to take away all guns, which is a distortion of the original argument.

Red Herring (RH): Diverting attention from the real issue by introducing an irrelevant topic.
Example: During a debate on environmental policies, a politician digresses to the opponent's personal life instead of addressing the policy issue.

Cherry Picking (CP): Selectively presenting evidence that supports one's argument while ignoring evidence that contradicts it.
Example: A report on climate change highlights only data supporting global warming, ignoring data that suggests otherwise.

False Analogy (FA): Making a misleading comparison between two things that are not truly comparable.
Example: Comparing the job of a teacher to that of a babysitter, implying they require similar skills and should therefore be compensated similarly.

Hasty Generalization (HG): Making a generalized statement based on a small or unrepresentative sample.
Example: Meeting three aggressive dogs and concluding that all dogs are aggressive.

Post Hoc (PH): If because one event followed another, the first event caused the second.
Example: Believing that carrying a lucky charm resulted in winning a game, just because the win came after starting to carry the charm.

False Cause (FC): Mistaking correlation for causation, if because two events occur together, one causes the other.
Example: Asserting that ice cream consumption causes drowning because both increase during the summer.

Vagueness (VAG): Using imprecise, unclear, or ambiguous language in an argument.
Example: A politician says they will "make the economy better" without specifying any actual policies or steps.





Please state all logical fallacies that are present in the text from this list. Explain where the fallacy occurs and why.  

Output a JSON (example):
\begin{verbatim}
{"cases": [{
"name": "Title",
"source": "WSJ",
“sentences”: {1: “sentence 1”,2: “sentence 2”}
"fallacies": {
    "logical_fallacies": ["CP", "RH"],
    "sentences": { 
        "CP": [4, 5, 10],
        "RH": [17, 18]},
    "annotations": {
        "CP": {
            "L1": [{
                "explanation": "insert explanation here",
                "sentence": [4,5], 
                “link”: “URL”}],
            "L2": [{
                "explanation": " insert explanation here ",
                "sentence": [4,5]}],
            "L3": [{
                "explanation": "",
                "sentence": [4,5]}]
        }
    }
}}]}
\end{verbatim}
Level 1 (L1) focuses on immediate correction, Level 2 (L2) provides a detailed analysis with evidence, and Level 3 (L3) aims to preemptively inform and educate the reader about potential misinformation. All explanations should be filled.

Each explanation should:
Be supported by specific examples and evidence from reputable sources. Incorporate at least one outbound link to Google or Bing that offers a contrasting viewpoint or additional data if appropriate. Please only use Google or Bing, replacing the query with a contrasting idea. For example, if you suggest looking at examples where “the earth is not flat” then provide the link https://www.google.com/search?q=the+earth+is+not+flat
Output in JSON code environment.

\clearpage
\section{Quantitative Study Results}
\label{appendix:study_results}

\begin{figure}[h]
\centering	
\includegraphics[width=0.95\linewidth]{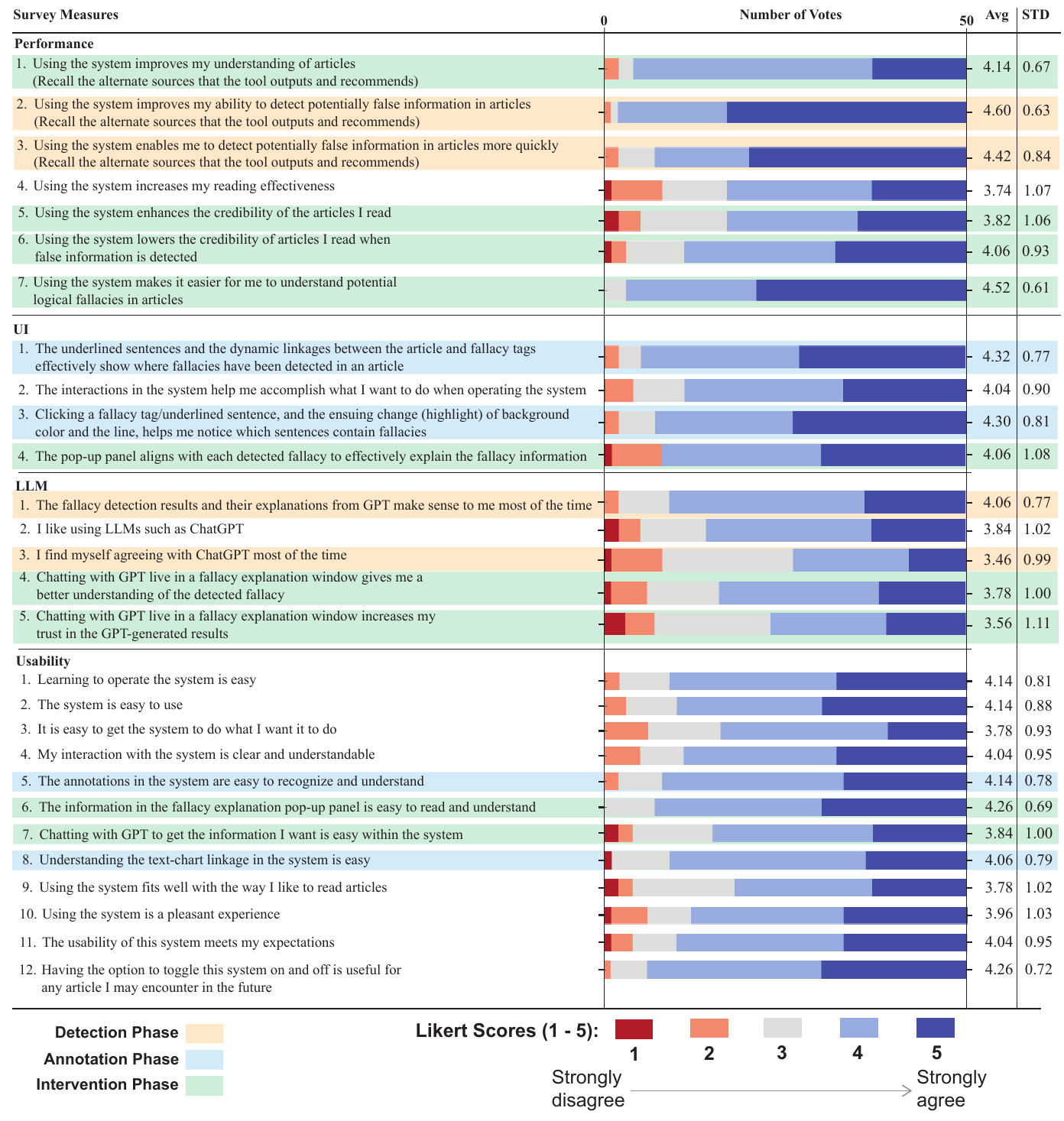}
\caption{Survey measures and the results of the crowdsourced study as described in Section~\ref{subsection:crowdsourced_study}. 
}
\Description{(Table) This figure shows the survey measure text and the study results.}
\label{fig:full_study_result}
\end{figure} 

\clearpage
\section{Additional Online News Cases and Detected Fallacies}
\label{appendix:case_appendix}

\begin{figure}[h]
\centering	
\includegraphics[width=0.85\linewidth]{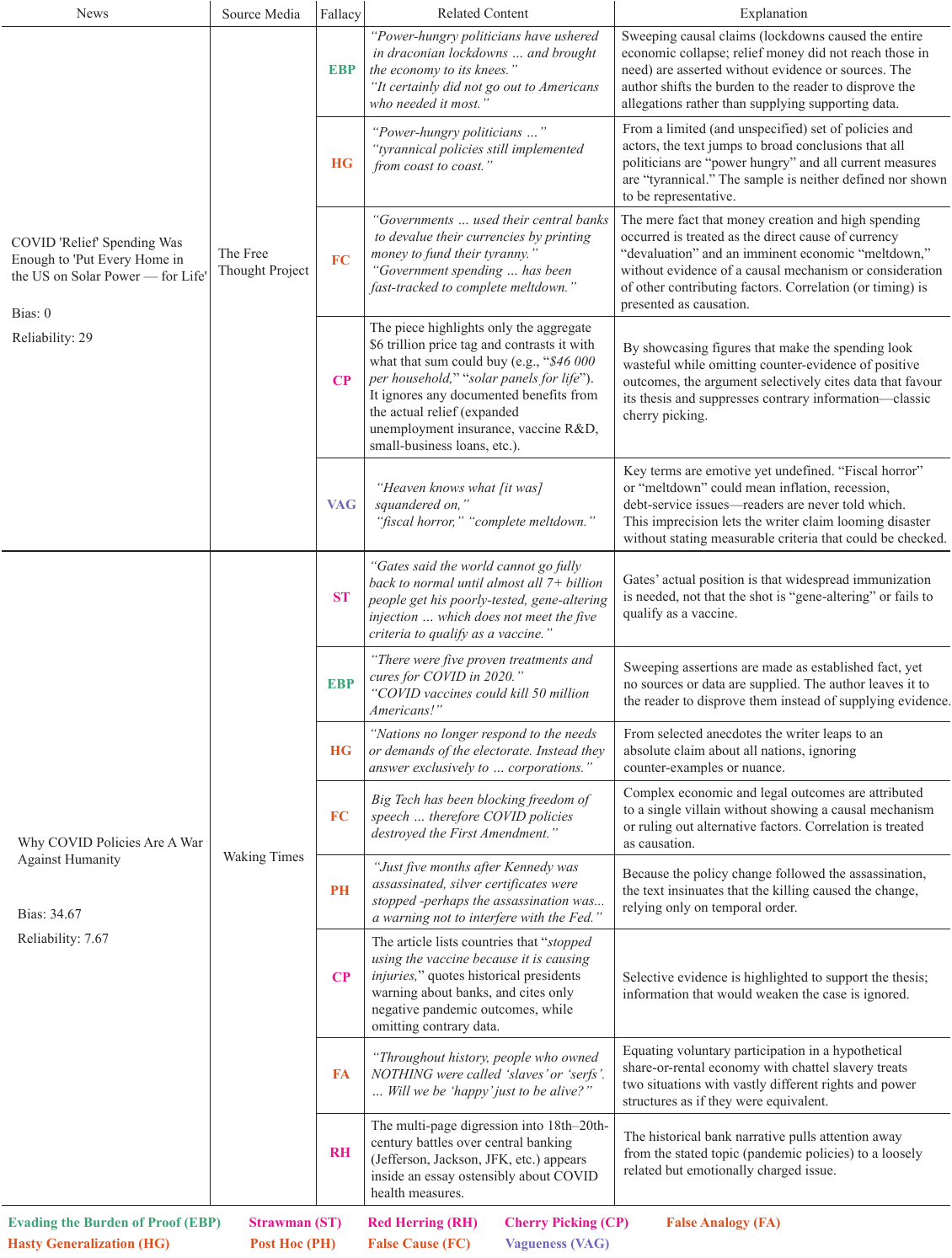}
\caption{\addition{Additional online news article cases in which the specified logical fallacies were detected.}}
\Description{(Table) This figure identifies the news cases with logical fallacies.}
\label{fig:case_appendix}
\end{figure}

\end{document}